\documentclass[12pt]{article}
\usepackage{epsf}
\newcommand{\F}{\Phi}
\newcommand{\HH}{{\cal H}}
\newcommand{\CN}{{\cal N}}
\newcommand{\G}{\Gamma}
\newcommand{\om}{\omega}
\newcommand{\<}{\langle}
\begin{document}
\date{}
\title{Time-Reversal and Irreversibility\thanks{
Based on three lectures delivered at the
``Second German-Polish Symposium on New Ideas in
the Theory of Fundamental Interactions'', Zakopane,
Poland September 1995.}}
\author{A.\ Bohm\\
{\small\sl Department of Physics, University of Texas, Austin, TX 78\,712,
U.S.A}\\
\and
P.\ Kielanowski\\
{\small\sl Institute of Theoretical Physics, Warsaw University, Poland}\\
{\small\sl and}\\
{\small\sl Departamento de F\'\i sica, CINVESTAV del IPN, Mexico}
 }
\maketitle
\begin{abstract}
The time reversal and irreversibility in conventional quantum mechanics
are compared with those of the rigged Hilbert space quantum mechanics.
We discuss the time evolution of Gamow and Gamow-Jordan vectors and
show that the rigged Hilbert space case admits a new kind of irreversibility
which does not appear in the conventional case. The origin of this
irreversibility can be traced back to different initial-boundary conditions
for the states and observables. It is shown that this irreversibility does
not contradict the experimentally tested consequences of the time-reversal
invariance of the conventional case but instead we have to introduce a new
time reversal operator.
\end{abstract}
\section{Introduction}
Irreversibility in the title refers to intrinsic irreversibility for
quantum physical systems on the microphysical level; this means there exist
microphysical ``states'' $\psi$ whose time evolution
(generated by an essentially self-adjoint semibounded Hamiltonian $H$)
$\psi(t)=\mbox{``e}^{-iHt/\hbar}\mbox{''}\psi(0)$ has a preferred direction,
$t\geq0$,~\cite{R1}. Time reversal in the title refers to the existence of an
operator $A_T$ which is usually viewed as associating to every state vector
$\phi(t)$ a state $\phi^\prime(-t)\equiv A_T^{-1}\phi(-t)$
at the negative time $-t$ (relative to a distinguished point of time $t=0$).
Irreversibility and time-reversibility thus appear to be in conflict with each
other. Here we want to discuss the resolution of this conflict, based on the
empirical fact, that (due to the boundary and initial conditions) for a given
state one can usually not (experimentally) prepare its time-reversed state, and
that the experimentally tested time-reversal invariance (like, e.g.,
reciprocity relations) refers to the relations of $A_T$ with the observables and
not the action of $A_T$ on
the states.

\section{Irreversibility, initial-boundary conditions, the
time-evolution semigroup and Gamow vectors
}
Standard (Hilbert Space) quantum mechanics admits only reversible time
evolution because time evolution is represented by a group (generated by a
self-adjoint Hamiltonian). In contrast to this mathematical theory, there
is ample empirical evidence of intrinsically irreversible time evolution
of microphysical systems, e.g., the decay of quasistationary states or
resonances.  Truly stationary states of such quantum physical systems like
stable elementary particles are rare.  Most relativistic or
non-relativistic elementary particles are decaying states (weakly or
electromagnetically) or hadron resonances.  Empirically, stability or the
values of the lifetime does not appear to be a criterion for
elementarity.  Stable particles are not qualitatively different from
quasistable particles, but only quantitatively different by a zero or
negligible value of the width $\Gamma$.  (A particle decays if it can decay and
it is stable if selection rules for some quantum numbers prevent it from
decaying.)

Resonances have a preferred direction of time (arrow of
time).  If one takes the point of view that resonances are autonomous
quantum-physical entities and decaying particles are {\it not} less
fundamental than stable particles, then one needs a mathematical theory
which includes semigroup time evolution. Further, if both stable and
quasistable states should be described on the same footing and, since
there are state vectors for stable states, there should also be state
vectors for quasistable states. The state vector of a resonance, however,
needs to have irreversible time evolution.

The standard way in which irreversibility is introduced in quantum theory
is through the master equation~\cite{R2}
\begin{equation}
  {\partial\rho(t)\over\partial t}
  =
  L \rho(t)
\end{equation}
%$$equation \: 1 = (2.1)$$
where $\rho (t)$ describes the state of the system $S$, the Liouville
operator $L$ is given, e.g., by~\cite{R2},~\cite{R3}
\begin{equation}
  L \rho(t)
  =
  -
  {i\over\hbar} \left[H,\rho(t)\right]
  +
  \delta H \varrho
\end{equation}
% $$equation \: 2 = (2.2) \:{\rm with \: second \: term \: on \: r.h.s.
% \: replaced \: by} \: \delta H \varrho $$
For $\delta H=0$, (1) with (2) is the irreversible time evolution of
the isolated quantum system (von Neumann equation).  The term $\delta H
\varrho$ represents some complicated external effects upon the
non-isolated system. With this term~(1) is the standard way of describing
extrinsic irreversibility due to the effect of an external reservoir $R$
(e.g., measuring apparatus) upon the system.
This irreversible time evolution is described by a semigroup
$\rho(t)=\Lambda(t)\rho(0)$
generated by the Liouvillian $L$,
$\Lambda(t)=\mbox{e}^{Lt}$, $t\geq0$.
Equation (1) has also
been applied to the time evolution of such microphysical systems as
the $K_L - K_S$ meson system~\cite{R3}.

That a fundamental concept like irreversibility should be caused by
extrinsic influences has been considered unsatisfactory by many people
working on irreversibility and statistical physics.  According to
Prigogine's ideas~\cite{R4}, irreversibility should be intrinsic to the
dynamics and should have its origin in the resonances (Poincar\'{e}
resonances) rather than being caused by merely external effects of a
quantum reservoir or the irreversible act of a measurement apparatus.
This requires also a dynamical semigroup
which, however, should be
generated by the Hamiltonian $H$, $\rho(t)=\mbox{e}^{-iHt}\rho(0)
\mbox{e}^{iHt}$, and not by a Liouvillian like~(2).

The idea of intrinsic irreversibility and the empirical facts
of resonances can both be accommodated by a new
mathematical theory which is similar to the standard (von Neumann) quantum
mechanics (nonrelativistic or relativistic) but uses a different
mathematical idealization~\cite{R11a}.

The interpretation of this new quantum theory is, like the Hilbert space
idealization, based on the Copenhagen interpretation of quantum mechanics,
but it makes a much more distinct separation between the {\it state} and
the {\it observable}.  The state is defined by a preparation apparatus that
prepares the state and is described mathematically by a statistical
operator (density matrix) or a state vector. The observable is defined by a
registration apparatus that measures its values in the state and is
mathematically described by self-adjoint operators and their projectors
(in place of the projection operator $\mid\psi\rangle\langle\psi\mid$
one can also take the vector $\psi$ up to a phase factor to describe
this observable).

The mathematical formulation of this new quantum theory uses also a linear
topological space, but instead of von Neumann's Hilbert space it uses the
Gelfand triplet (also called rigged Hilbert space (RHS)).

Rigging the Hilbert space may turn many people away from this subject because it
may appear to some as an unnecessary mathematical complication (or even a
disreputable practice).

This is really not the case, because on the level of the mathematical rigor
employed by the physicist the RHS formulation of quantum mechanics is like
Dirac's bra-
and ket-formalism. When physicists talk about the Hilbert space they mostly mean
a pre-Hilbert space, i.e., a linear space $\Psi$ with a ``scalar product'',
denoted by $(\psi,F)$ or $\langle\psi\mid F\rangle$ without worrying about its
topological completion.
The Hilbert space of mathematicians is a much more complicated structure, its
elements being represented not by functions but by classes of functions whose
elements differ on a set of Lebesgue measure zero, a mathematically complicated
and physically useless concept (because the apparatus resolution is described
by smooth functions). The RHS is the same linear space $\Psi$ only with
different topological completions: one completes $\Psi$ with respect to a
topology that is stronger than the topology given by the ${\cal H}$-space norm
(e.g., one uses a countable number of norms) to obtain the space
$\Phi\subset{\cal H}$ and considers in addition the topological dual to $\Phi$,
i.e., the space of {\em continuous} antilinear functionals of $\Phi$ denoted by
$\Phi^\times$. Then one obtains the triplet
\begin{equation}
\begin{array}{rccc}
\mbox{Gelfand triplet: }\Psi\subset
&\Phi&\subset{\cal H}={\cal H}^\times\subset &\Phi^\times\\
&\in &&\in\\
\mbox{with elements ``bra'' and ``ket''\,\,\,\,}&\langle\phi\mid
&&\mid F\rangle\\
\mbox{or ``ket'' and ``bra''\,\,\,\,}
&\mid\phi\rangle &&\langle F\mid
\end{array}
\label{3a}
\end{equation}
A widespread example for $\Phi$ is the Schwartz space.

The vectors $\phi\in\Phi$
(in their form as kets $\mid\phi\rangle$ or bras $\langle\phi\mid$)
represent physical quantities connected with the experimental apparatuses (e.g.,
state $\phi$ defined by a preparation apparatus or an observable
$\mid\psi\rangle\langle\psi\mid$ defined by a registration
apparatus (detector)
fulfill $\phi,\psi\in\Phi$), the vectors $\langle F\mid$ or $\mid
F\rangle\in\Phi^\times$ represent quantities connected with the microphysical
system (e.g., ``scattering states'' $\mid E\rangle$ or decaying states $\mid
E-i\Gamma/2\rangle$). ${\cal H}$ itself does not have any special physical
meaning.

A general observable is now represented by a bounded operator $A$ in $\Phi$
(but in general
by an unbounded $\overline{A}$ or $A^\dagger$ in ${\cal H}$)
and corresponding to the triplet~(\ref{3a})
one has now a triplet of operators
\begin{equation}
A^\dagger\left|_\Phi\right.\subset A^\dagger\subset A^\times
\label{3b}
\end{equation}
In here $A^\dagger$ is the Hilbert space adjoint of $A$
(if $A$ is essentially self adjoint then $A^\dagger=\overline{A}$),
$A^\dagger\left|_\Phi\right.$ denotes its restriction to the space $\Phi$,
and the operator $A^\times$ in $\Phi^\times$
is the conjugate operator of $A$ defined by
\begin{equation}
\langle A\phi\mid F\rangle=
\langle \phi\mid A^\times F\rangle\,\,\,\,
\mbox{for all $\phi\in\Phi$ and all $\mid F\rangle\in \Phi^\times$}.
\label{3c}
\end{equation}
By this definition $A^\times$ is the extension of the operator
$A^\dagger$ to the space $\Phi^\times$ (and not the extension of the operator
$A$ which is most often used in mathematics).
A very important point is that the operator $A^\times$ is only defined for an
operator $A$ which is continuous=bounded in $\Phi$, then $A^\times$ is a
continuous (but not bounded) operator in $\Phi^\times$.
It is impossible in quantum mechanics (empirically)
to restrict oneself to continuous=bounded operators $\overline{A}$
in ${\cal H}$, but one can restrict oneself to algebras of observables
$\{A,B\ldots\}$ described by continuous operators in $\Phi$.
Then $A^\times,B^\times\ldots$ are defined and continuous in $\Phi^\times$.
If $A$ in~(\ref{3c}) is not self-adjoint then $A^\dagger\mid_{\Phi}$
need not be a continuous operator in $\Phi$ even if $A$ is,
but one can still define the conjugate $A^\times$
which is continuous in $\Phi^\times$.

A generalized eigenvector $F\in\Phi^\times$ of an operator $A$ is defined by
\begin{equation}
\langle A\phi\mid F\rangle=
\langle \phi\mid A^\times F\rangle=
\omega\langle \phi\mid F\rangle\,\,\,
\mbox{for all $\phi\in\Phi$}
\label{3d}
\end{equation}
where the complex number $\omega$ is called the generalized eigenvalue.
This is also written as
\begin{equation}
A^\times\mid F\rangle=
\omega\mid F\rangle.
\label{3e}
\end{equation}
For an essentally self-adjoint operator
$A^\dagger=\overline{A}$ (= closure of $A$)
this is often also written as
\begin{equation}
A\mid F\rangle=
\omega\mid F\rangle
\label{3en}
\end{equation}
especially if one suppresses the mathematical subtleties and acts as if one has
just a linear scalar product space $\Psi$.

Calculating just in the pre-Hilbert space $\Psi$ --- as physicists usually do
--- the RHS formulation is really not more difficult than the Hilbert space
formulation. One just has to use a slightly more general set of rules for these
calculations. This has always been done in the Dirac formalism of bra's and
ket's. In addition to the rules of the Dirac formalism, the RHS provides a
mathematical justification for additional rules of
mathematical manipulations.
The most
important of these are:
\begin{enumerate}
\item
the eigenvectors of self-adjoint observables $A$ (i.e.\ with
$A^\dagger=\overline{A}$) in~(\ref{3en})
can be complex
\item
the time evolution for some of the solutions of the
Schr\"odinger equation can be given by a semigroup and not by a reversible
unitary group
\item
some vectors can decay exponentially (as envisioned by
Gamow).
\end{enumerate}

Dynamical equations
(laws of nature) are the same in both the Hilbert space and the RHS
formulations, namely given by the
Schr\"{o}dinger equation
\begin{equation}
  i\hbar {\partial\left|\phi(t)\right\rangle\over\partial t}
  =
  H \left|\phi(t)\right\rangle.
\end{equation}
or the von Neumann equation (2) with $\delta H=0$.
But in the RHS formulation
different initial and boundary conditions than in the Hilbert space
formulation allow for a greater variety of solutions; (this goes back to
Dirac (kets $\mid E \rangle$), Gamow (exponentially decaying ``state"
vectors
$\mid E - i \Gamma/2 \rangle$) and Peierls (purely outgoing boundary
conditions). These new vectors are in the rigged Hilbert space, $\mid E
\rangle, \mid E - i \Gamma/2 \rangle\, \in \Phi^{\times} \supset {\cal H}
\supset \Phi$, but not in the Hilbert space $\cal H$. Distinct
initial-boundary conditions for state vectors (e.g., {\em in}-states
$\phi^{+}$ of a scattering experiment) and observables $| \psi^-\rangle
\<\psi^- |$ (e.g., so-called {\em out}-states $\psi^-$ of a scattering
experiment) lead to two different rigged Hilbert spaces~\cite{R1},
whose precise mathematical properties had been defined earlier~\cite{R24}:
\begin{eqnarray}
&\F_-\subset\HH\subset\F^\times_-,\qquad\qquad&
\mbox{for ensembles or states}\label{G4a}\\
&\F_+\subset\HH\subset\F^\times_+,\qquad\qquad&
\mbox{for observables or effects}
\label{G4b}
\end{eqnarray}
The Hilbert space ${\cal H}$
is the same in both RHS's~(\ref{G4a}) and~(\ref{G4b})
and $\Phi$ of~(\ref{3a}) is $\Phi=\Phi_-+\Phi_+$ with
$\Phi_-\cap\Phi_+\neq\emptyset$.

In~(\ref{G4a}), $\Phi_-$ describes the possible state
vectors (preparation apparatus, e.g.,
$\phi^{\mbox{\rm\scriptsize in}}$ or $\phi^{+}$ of a scattering experiment)
and
$\Phi_{+}$ in~(\ref{G4b})
describes the possible observables  (e.g.,
$| \psi^{\mbox{\rm\scriptsize out}}\rangle
\<\psi^{\mbox{\rm\scriptsize out}} |$ or
$|\psi^-\rangle\<\psi^- |$ of a scattering experiment).

For the typical scattering experiment the physical meaning of $\Phi_-\ni\phi^+$
is depicted in Fig.\ 1. The {\em in}-state $\phi^+$ (precisely the state which
evolves from the prepared {\em in}-state $\phi^{\mbox{\rm\scriptsize in}}$
outside the interaction region where $V=H-H_0$ is zero) is determined by the
accelerator. The so called {\em out}-state $\psi^-$ (or
$\psi^{\mbox{\rm\scriptsize out}}$) is determined by the detector;
$\mid\psi^{\mbox{\rm\scriptsize out}}\rangle\langle
\psi^{\mbox{\rm\scriptsize out}}\mid$
is therefore the observable which the detector registers and not a state.
In the conventional formulation one describes both
the
$\phi^{\mbox{\rm\scriptsize in}}$ and the $\psi^{\mbox{\rm\scriptsize out}}$
by any vectors of the Hilbert space. In reality the
$\phi^{\mbox{\rm\scriptsize in}}$
(and $\phi^+$) and
$\psi^{\mbox{\rm\scriptsize out}}$
(and $\psi^-$) are subject to different initial and boundary conditions
and should therefore be described by different sets of vectors.
\begin{figure}
\epsfxsize=\textwidth
\epsffile{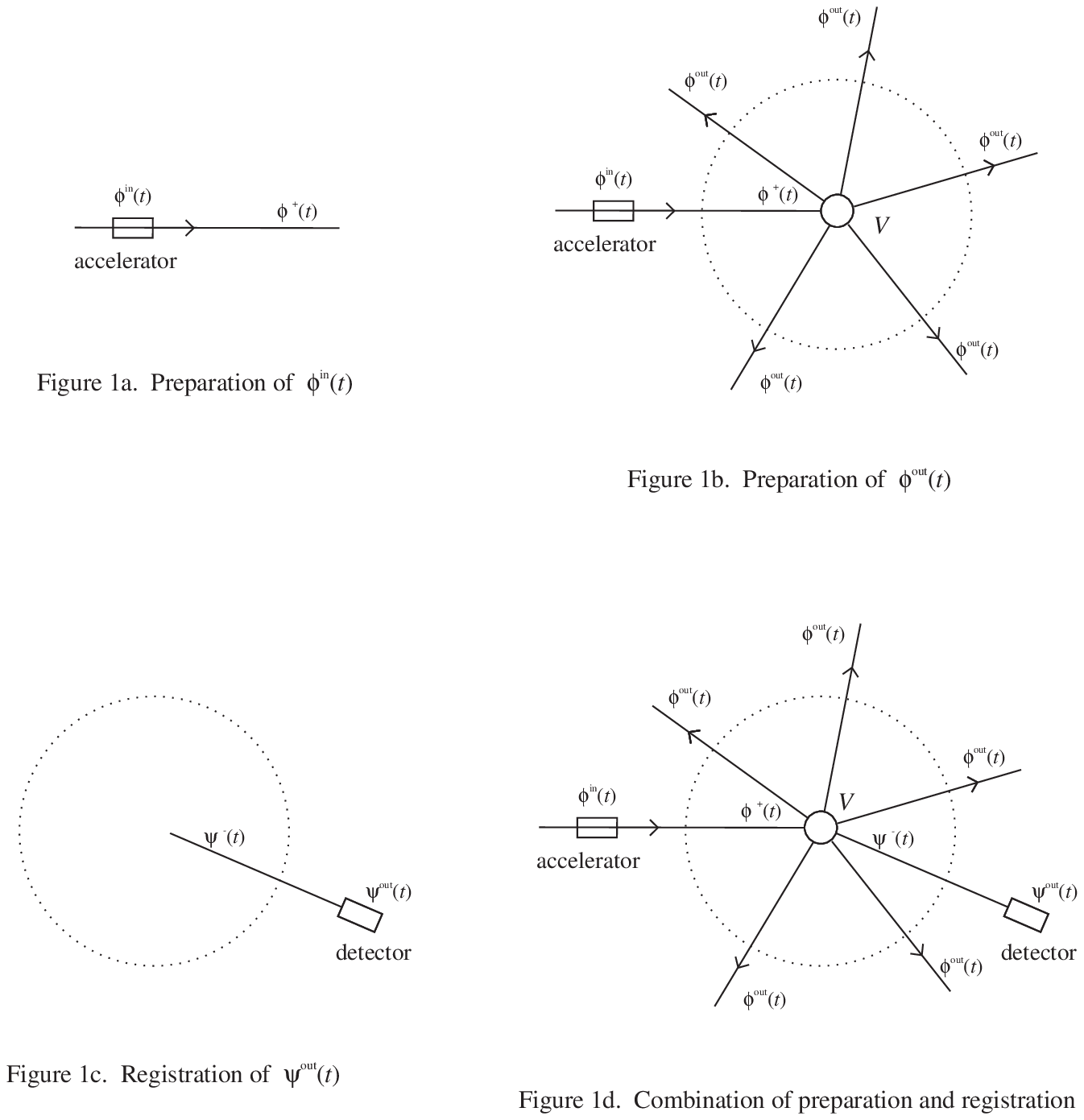}
\end{figure}

These distinct
initial-boundary conditions for state vectors and observable vectors
are stated as an
``arrow of time" in the form~\cite{R1}:
\begin{equation}
\mbox{
\begin{minipage}{0.8\textwidth}
  The state $\phi (t) \in \Phi_-$ must be prepared before an
observable $| \psi \rangle\<\psi |$ (with $\psi \in \Phi_{+}$) can be
measured in that state; i.e., if $t=0$ is the time before which the
preparation is completed and after which the registration begins, then
$\phi (t)$ must be prepared by a time $t \leq 0$.
\label{G5}
\end{minipage}
}
\end{equation}
The property of spaces in~(\ref{G4a}) and~(\ref{G4b}) can be derived from a
mathematical formulation of the ``arrow of time"~(\ref{G5})
using the Paley-Wiener
theorem~\cite{R1}. It turns out that  $\Phi_-$ is the
space of well behaved Hardy class vectors from below and $\Phi_{+}$
is the space of well behaved Hardy class vectors from above~\cite{R43}.
These are the same mathematical properties that had been obtained
earlier~\cite{R24},~\cite{R11} from the existence conditions for Gamow vectors.
The notation
$\phi^{+} \in \Phi_-$ and $\psi^- \in \Phi_{+}$
of opposite
sub- and superscripts for vectors and spaces
has no significance but is just a
consequence of the fact that the nomenclature in physics (scattering
theory for $\phi^+$, $\psi^-$) and mathematics (theory of Hardy class functions
for $\Phi_-$ and $\Phi_+$)
had been developed independently.

The semi-group of time evolution, and therewith irreversibility on the
microphysical level, is a mathematical consequence of the bi-partition of
the rigged Hilbert space into the two rigged Hilbert spaces~(\ref{G4a})
and~(\ref{G4b}) and
therewith of the dichotomy of state and observables and their ``arrow
of time"~(\ref{G5}).

In conventional quantum mechanics in Hilbert space the time evolution of
a state
\begin{equation}
 W (t)
  =
  U^\dagger(t) W (0) U(t)
  =
  e^{-iHt/\hbar} W (0) e^{iHt/\hbar},
  \hskip 10mm
  -\infty < t < \infty.
  \label{G6}
\end{equation}
is given by a group
\begin{equation}
 U^{\dagger} (t) = e^{-iHt/\hbar} \:\:\:\:\:\: -\infty
<t < + \infty
\end{equation}
Therefore for every statistical operator (or density matrix)
$W(t)$ one obtains (by calculation) a state operator
$W^{\mbox{\rm\scriptsize neg}} (t) \equiv W(-t)$.

In the rigged Hilbert
spaces~(\ref{G4a}),~(\ref{G4b})
we have the two extensions of the Hilbert space operator
$U^{\dagger}(t)$:
\begin{eqnarray}
&\mbox{the conjugate of}\,\,\,
U\mid_{\Phi_-}:\,\, U^{\dagger}(t)
\subset \: U_-^{\times} = e^{-iH^{\times}t/\hbar};\,\,\,
\mbox{for $t\leq0$}
\label{G8+}\\
&\mbox{the conjugate of}\,\,\,
U\mid_{\Phi_+}:\,\, U^{\dagger}(t)
\subset \: U_+^{\times} = e^{-iH^{\times}t/\hbar};\,\,\,
\mbox{for $t\geq0$}
\label{G8}
\end{eqnarray}
where $U^{\times}_{\pm}$ ($H^{\times}$)
denote the extensions of the unitary (self-adjoint) operators
$U^{\dagger} (t)$ ($H^{\dagger}=H$) to the spaces $\Phi^{\times}_{\pm}$.
It turns out that, mathematically, $U_-^{\times}$ in $\Phi^{\times}_-$
can only be defined by~(\ref{3c})  for values of the parameters $t \leq 0$,
 since for $t>0$ $U$ is not continuous in $\Phi_-$.
By the same arguments
$U^{\times}_{+}$ in
$\Phi^{\times}_{+}$ can only be defined for values of the parameters
$t \geq 0$.  This is the mathematical strategy by which the semigroup time
evolution is obtained. In the physical
interpretation of the mathematical  theory it is based on the ``arrow of
time"~(\ref{G5}). Now
one can no more define for every state $| \phi^-(t)\rangle\<\phi^-(t)|$, $\phi^-
\subset \Phi_{+}$ a state $W^{\mbox{\rm\scriptsize neg}} (t) \equiv |
\phi^-(-t)\rangle\<\phi^-(-t)|$, which seems to reflect the experimental
situation better (time reversal transforms
{\em out}-states of scattering experiment
which are highly correlated spherical waves, into highly correlated
incoming spherical waves
that go into
outgoing uncorrelated plane waves).
But an experiment in which highly correlated incoming spherical
waves go into uncorrelated plane waves is practically impossible to set up.

Summarizing, if one wants an irreversible time evolution on the microphysical
level one needs a mathematical idealization (i.e., a topological completion of
the linear algebraic space) which uses not Hilbert space
but the rigged Hilbert space. This
quantum theory in rigged Hilbert space has the following properties:
\begin{enumerate}
\item[{I.}] It has Dirac kets (scattering states) $\mid E\rangle$ and an algebra
of observables.
\item[{II.}] It has vectors, called Gamow vectors which we also denote by
kets as
$|\psi^G \rangle = |z^-_R \rangle \sqrt{2\pi \Gamma}$, that have the following
properties which make them ideally suited for the description of resonance
states in quantum theory:
%\end{enumerate}
\begin{enumerate}
\item[1.]  They are generalized eigenvectors of Hamiltonians $H$ (which we
always assume to be (essentially) self-adjoint and bounded from below)
with generalized eigenvalues $z_R = E_R - i \Gamma /2$,
\begin{equation}
H^\times  \left|\psi^{G}\right\rangle=z_{R}\left|\psi^{G}\right\rangle
\label{G9}
\end{equation}
where $E_R$ and $\Gamma$ are respectively interpreted as the resonance
energy and width.
\item[2.]  They satisfy the following exponential decay law for $t \geq 0$
only:
\begin{eqnarray}
  &W^{G}(t)
  &=
  \mbox{``}\, e^{-iHt/\hbar} \,\mbox{''}
  \left|\psi^{G}\right\rangle
  \left\langle\psi^{G}\right|
  \mbox{``}\, e^{iHt/\hbar} \,\mbox{''}\nonumber\\
  &&=
  e^{-i\left(E_{R} - i\,{\Gamma /2}\right)t/\hbar}
  \left|\psi^{G}\right\rangle
  \left\langle\psi^{G}\right|
  e^{i\left(E_{R} + i\,{\Gamma /2}\right)t/\hbar}\nonumber\\
  &&=
  e^{-\Gamma t/\hbar}\,
  W^{G}(0)
  \label{G10}
\end{eqnarray}
\item[3.]  They have a Breit-Wigner energy distribution.
\item[4.]  They obey an exact Golden Rule of which Fermi's Golden Rule is the
Born approximation.
\item[5.]  They are associated with a pole at $z_R$ in the second sheet of the
analytically continued $S$-matrix.  They are derived as the functionals of
the pole term of the $S$-matrix.
\end{enumerate}
\end{enumerate}

In the absence of a vector description of resonances in the Hilbert space
formulation, the pole of the $S$-matrix has commonly been taken as the
definition
of a resonance.
Since in the RHS formulation the Gamow vectors are derived from the pole term of
the $S$-matrix~\cite{R11}, these vectors $\mid z_R^-\rangle\in\Phi_+^\times$
define decaying resonance states as autonomous microphysical entities.
 (There are also Gamow vectors $|{z^*_R}^+\rangle$, $z^*_R = E_R + i
\Gamma/2$  associated with the pole at $z^*_R$, which have an exponentially
growing semi-group evolution for $- \infty <t \leq 0$).

\section{Gamow-Jordan vectors --- a mathematical actuality and a physical
%possibility.\cite{R11a}}
possibility.}

The mathematical definition of Dirac kets was given in 1966~\cite{R9}, the
Gamow
vectors were introduced about 1976~\cite{R10},\cite{R11};
a generalization of Gamow vectors to
higher order poles of the $S$-matrix was given in 1995~\cite{R12}. An ${\cal
N}$-th
order $S$-matrix pole at the complex energy $z_\CN=E_\CN-i\gamma_\CN$ has $\CN$
Gamow vectors of order $0,1,\ldots k\ldots (\CN-1)$:
\begin{equation}
|z_\CN^-\rangle^{(0)},\,\,
|z_\CN^-\rangle^{(1)},{\ldots,}|z_\CN^-\rangle^{(k)},
{\ldots,}|z_\CN^-\rangle^{(\CN-1)}
\label{G11}
\end{equation}
associated with it. The $k$-th order Gamow vector $|z_\CN^-\rangle^{(k)}$
is a Jordan
vector of degree $(k+1)$, i.e. it fulfills the eigenvalue equations~\cite{R59}
\begin{eqnarray}
&&({H}^\times-z)^k|z_\CN^-\rangle^{(k)}=0;\nonumber\\
&&{H}^\times|z_\CN^-\rangle^{(k)}=z_\CN|z_\CN^-\rangle^{(k)}
+|z_\CN^-\rangle^{(k-1)}\,\,\,\,\mbox{for}\,\,\,\,
k=0,1,\ldots,(\CN-1)
\end{eqnarray}
These equations are, like the eigenvector equation for Dirac kets and for Gamow
vectors (= Gamow vectors of order 0 = Jordan vectors of degree 1), understood as
generalized eigenvector equations (i.e., functionals) over the space $\Phi_+$:
\begin{eqnarray}
\<{H}\psi^-|z^-_{\CN}\rangle^{(k)}
&\equiv&\<\psi^-|{H}^\times|z^-_{\CN}\rangle^{(k)}\nonumber\\
&=&z_{\CN}\<\psi^-|z^-_{\CN}\rangle^{(k)}
+\langle\psi^-\mid z^-_{\CN}\rangle^{(k-1)}
\,\,\,\,\mbox{for all}\,\,\psi_-\in\Phi_+.
\label{G12}
\end{eqnarray}
This means $|z_\CN^-\rangle^{(k)}\in\Phi_+^\times$, and the
$\CN$-th order $S$-matrix pole is associated to a
$\CN$-dimensional subspace
${\cal M}_{z_{\CN}}\subset\Phi_+^\times$,
spanned by the $|z_\CN^-\rangle^{(k)}$, $k=0,1,\ldots,(\CN-1)$,
i.e., to the set of all
\begin{equation}
|z_\CN^-\}=\sum_{k=0}^{\CN-1}|z_\CN^-\rangle^{(k)}c_k,
\,\,\,\,c_k\in{\cal C}.
\end{equation}
On ${\cal M}_{z_\CN}\subset\Phi_+^\times$
the Hamiltonian ${H}^\times$ (i.e., the extension of the self-adjoint
operator ${H}^\dagger$ to $\Phi^\times$) is not diagonalizable, but can only
be brought into the normal form of a Jordan block:
\begin{equation}
H^\times_\CN \Longleftrightarrow\left(
\begin{array}{cccccc}
              z_\CN&0&\cdots&\cdots&\cdots&0\\
              1&z_\CN&&&&\vdots\\
              0&1&\ddots&&&\vdots \\
              \,\vdots&&\ddots&\ddots&&\vdots\\
              \,\vdots&&&\ddots&\ddots&0\\
              0&0&\cdots&\cdots&1&z_\CN
\end{array}
\right)
\end{equation}

From this follows that the matrix representation of the time evolution
operator~(\ref{G8}) on the ${\cal N}$-dimensional eigenspace
${\cal M}_{z_{\cal N}}$ is given by
\begin{equation}
\begin{minipage}{0.9\textwidth}
\epsfxsize=\textwidth
\epsffile{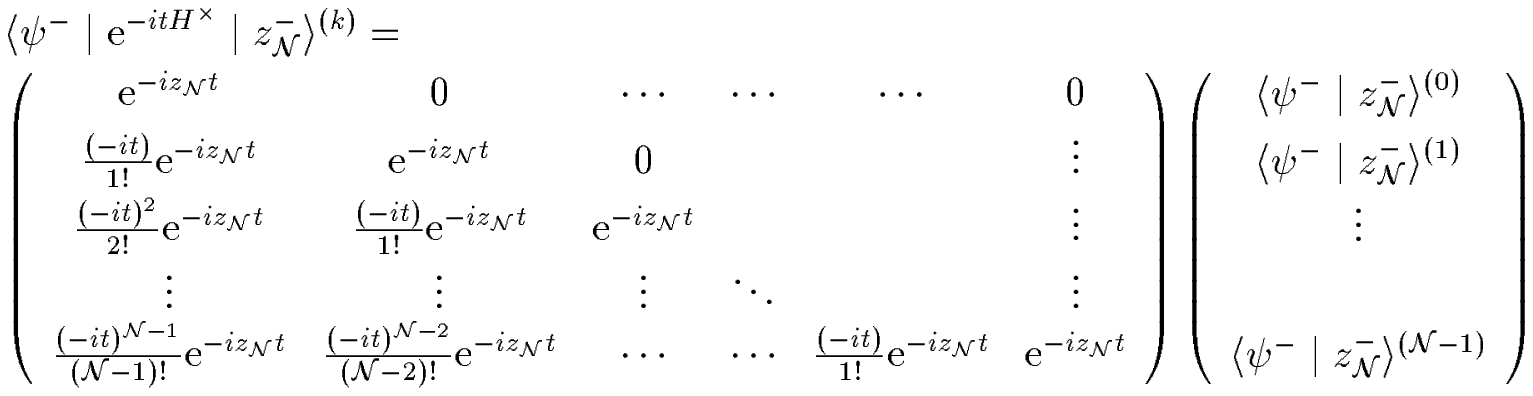}
\end{minipage}
\label{ar22-1}
\end{equation}
which can also be written as
\begin{eqnarray}
&&\mbox{e}^{-itH^\times}\mid z^-_\CN\rangle^{(k)}=
\mbox{e}^{-itz_\CN}\left\{\mid z^+_\CN\rangle^{(k)}
+\frac{(-it)}{1!}\mid z^+_\CN\rangle^{(k-1)}+\cdots\right.
\nonumber\\
&&+\left.
\vphantom{\mbox{e}^{-itH^\times}\mid z^-_\CN\rangle^{(k)}=
\mbox{e}^{-tz_\CN}\left\{\mid z^+_\CN\rangle^{(k)}
+\frac{(-it)}{1!}\mid z^+_\CN\rangle^{(k-1)}+\cdots\right.
}
\frac{(-it)^l}{l!}\mid z^+_\CN\rangle^{(k-l)}+
\cdots
+\frac{(-it)^k}{k!}\mid z^+_\CN\rangle^{(0)}
\right\}
\,\,\,\,\mbox{for $t\geq0$ only}.
\label{ar22-2}
\end{eqnarray}
This means that whereas the zeroth order Gamow state only decays
exponentially with time, the $k$-th order Gamow vector
$\mid z^-_\CN\rangle^{(k)}$ evolves into a superposition
of lower order Gamow vectors. After a long time (relative to the scale
set by $\hbar/\Gamma$) the most significant term is the zeroth
order Gamow vector $\mid z_\CN^-\rangle^{(0)}$,
whose time dependence is given by $\mbox{e}^{-itz_\CN}t^{k}$.

\section{The complex basis vector expansion and some of its consequences in
physical applications}

The most important result of the new mathematical theory of quantum physics in
the rigged Hilbert space is the complex eigenvector expansion.
This is the generalization of the elementary
basis vector
expansion of a
3-dimensional vector,
${\bf x}=\sum_{i=1,2,3}{\bf e}_i({\bf e}_i\cdot{\bf x})=\sum{\bf e}_i\cdot x_i$
to the expansion of vectors
$\phi^+\in\Phi_-$ using as basis vectors the generalized eigenvectors
$|z_{R_i}^-\rangle$ and $|z^-\!\rangle^{(k)}$ of
self-adjoint operators ${H}$ with
complex eigenvalues $z_{R_i}$, and $z$, respectively.

Earlier developments towards this generalization
were the fundamental theorem of linear algebra
which states that for every self-adjoint operator $H$ in a $n$-dimensional
Euclidian space ${\cal H}_n$ there exists an orthonormal basis $e_i\ldots e_n$
in ${\cal H}_n$ of eigenvectors $He_i=E_ie_i$. I.e., $f\in{\cal H}_n$
can be written
$f=\sum_{i=1}^ne_i(e_i,f)$.
This theorem generalizes to the infinite dimensional Hilbert space
${\cal H}$,  but only for self-adjoint operators $H$ which are completely
continuous (also called compact operators which include Hilbert-Schmidt,
nuclear, traceclass operators).
For an arbitrary self-adjoint operators $H$ one has to go outside the space to
find a complete basis system of eigenvectors (generalized).

 The {\em first} step
in this direction is the Dirac basis vector expansion which in mathematical
terms is called the nuclear spectral theorem. It states that for every
$\phi\in\Phi$
\begin{equation}
\phi=\int^{+\infty}_0\,dE\;|E^+\rangle\<^+E|\phi^+ \rangle\;+\;
  \sum_n\,|E_n)(E_n|\,\phi\, )\quad
\mbox{for }\;\phi \in\F
\end{equation}
In here, $|E_n)$ are the discrete eigenvectors of the exact
Hamiltonian $H=K+V$, (describing the bound states) $H|E_n)=E_n|E_n)$,
and $|E^+\rangle$ are the generalized eigenvectors (Dirac kets)
of $H$ corresponding
to the continuous spectrum (describing scattering states).  The
integration extends over the spectrum of $H$: $0\le E <\infty$; and
in place of the $|E^+\rangle$ one could also have chosen the $|E^-\rangle$,
if the {\em out}-wavefunctions are more readily available.

The {\em second} step is the ``complex basis vector expansion''
For every $\phi^+\in\Phi_-$ (a similar expansion holds also for
every $\psi^-\in\Phi_+$) one obtains for the case of
a finite number of resonances poles at the positions
$z_{R_i}\,,\;i=1,\;2,\cdots N$, the following basis
system expansion:
\begin{eqnarray}
\phi^+&=&
\int_0^{-\infty_{II}}d\om\;|\om^+\rangle\<^+\om|\phi^+\rangle\;+\;
\sum_{i=1}^N\,|z_{R_i}^-\rangle\;2\pi\G_i \;\<^+z_{R_i}|\phi^+\rangle\nonumber\\
&+&\sum_n |E_n)(E_n|\phi)\,\,\,\,
\mbox{for}\,\,\,\, \phi^+\in\F_-
\label{G16}
\end{eqnarray}
where
$|z_{R_i}^-\rangle\sqrt{2\pi\Gamma_i}=\psi^{G_i}\in\Phi^\times_+$
are Gamow kets~(\ref{G9}) representing decaying states~(\ref{G10}).

If we assume that there are two decaying
states $R_1 =S$ and $R_2 =L$ and no bound states,
then the pure state (prepared by the
experimental apparatus) has according to~(\ref{G16})
the following representation
in terms of the Gamow vectors $\psi_L^G=\mid z_L^-\rangle\sqrt{2\pi\Gamma_L}$,
$\psi_S^G=\mid z_S^-\rangle\sqrt{2\pi\Gamma_S}$
and the remaining part which we call $\phi^+_{\mbox{\rm\scriptsize bg}}$:
\begin{equation}
\phi^+=\psi^G_L\,b_L\;+\;\psi^G_S\,b_S\;+\;\int^{-\infty_{II}}_0 dE\;
|E^+\rangle\<^+E|\phi^+\rangle.
\label{N25}
\end{equation}
In here $b_L$ and $b_S$ are some complex numbers that depend upon
the ``normalization'' of the Gamow vectors $\psi^G_{L,S}$
(and of $\phi^+$), and upon some phase convention.  All the
vectors in the generalized basis system expansion are (generalized)
eigenvectors of the exact Hamiltonian, and, in particular, the
Gamow vectors $\psi^G_{L,S}$ are eigenvectors of
the exact Hamiltonian $H$, with complex eigenvalue
$\left( E_L -i\,{\G_L / 2} \right)$ and
$\left( E_S-i\,{\G_S / 2}\right)$, respectively.

We now apply
the time evolution operator to equation~(\ref{N25}). Since the
$\psi_i^G$ are elements of $\Phi^\times_+$
we can only apply the operator $U^\times_+(t)$ of~(\ref{G8})
to it and we obtain:
\begin{equation}
\phi^+ (t)\equiv e^{-iH^\times t}\phi^+=
   e^{-i(E_L-i\,\G_L /2)t}\,\psi^G_Lb_L\; +\;
e^{-i(E_S-i\,\G_S /2)t}\,\psi^G_Sb_s\; +\;\phi^+_{\mbox{\rm\scriptsize bg}}
(t)\; ; t\ge 0
\label{G19}
\end{equation}
Since the time evolution semigroup~(\ref{G8}) has the restriction $t\ge
0$, the same restriction must be used for~(\ref{G19}).
$\phi^+_{\mbox{\rm\scriptsize bg}} (t)$
is the time evolved background term
\begin{equation}
\phi^+_{\mbox{\rm\scriptsize bg}} (t)\equiv
\int^{-\infty_{II}}_{0} dE\; e^{-iEt}\,
|E^+\rangle\<^+E|\phi^+\rangle \;.
\end{equation}
These equations are understood as a functional equation
over all $\psi^-\in\F_+$. This means that
$\phi^+ (t)\in\Phi_-\subset\Phi_+^\times $ can be used to obtain
$\<\psi^-|\phi^+ (t)\rangle$,
whose modulus square is the probability to find the time evolved state by a
detector that detects the  observable $|\psi^-\rangle\<\psi^-|$ for
any $\psi^-\in\F_+$, but {\em not} for a $\psi^-\in\F_-$.

The result~(\ref{G19}) means that the time evolution of a superposition
of two (or more) Gamow states does not regenerate one Gamow state
from the other, or from the background $\phi^+_{\mbox{\rm\scriptsize bg}} (t)$.
In particular, if the state $\phi^+$ can be prepared such that
at some time $t_0 \ge 0$ the background term $\phi^+_{\mbox{\rm\scriptsize bg}}
(t)$ is practically zero, then it will remain zero for all $t >t_0$,
and the two Gamow states will evolve separately with their
separate exponential laws without regenerating each other:
\begin{equation}
\phi^+ (t)\approx e^{-iE_L t}\,e^{-(\G_L /2)t}\,\psi^G_L\,b_L \; +\;
e^{-iE_S t}\,e^{-(\G_S /2)t}\,\psi^G_S\,b_S
\label{G21}
\end{equation}
Approximations like~(\ref{G21}) have been used for the time evolution of
a two-resonance system (like the $K_L$--$K_S$-system with $\phi^+(t)$
representing the $K^0$ state~\cite{R13}) in theories with ``effective
Hamiltonians''
given by $2\times2$ complex diagonizable matrix.
These effective theories are usually
legitimized by  the
Wigner-Weisskopf approximation~\cite{R14}.
In our irreversible quantum theory
the expression~(\ref{G19})
is exact and it justifies to some extent the effective theory(\ref{G21}).
And~(\ref{G19}) shows that
the problem
of ``deviation from the exponential decay law''
or ``vacuum regeneration of $K_S$ from $K_L$''~\cite{R19} arises from
the artifacts of the Hilbert space mathematics
and can be overcome in the exact theory using the rigged Hilbert space.

However there is an extra term in~(\ref{G19}) which we called
$\phi_{\mbox{\rm\scriptsize bg}}^+$
and which is not taken into consideration in any of the finite
dimensional effective theories of complex Hamiltonians, in particular
not in the Lee, Oehme, Yang theory~\cite{R13} of the neutral Kaon system.
This term, which comes from the integral along the negative
real axis in the second sheet of the $S$-matrix, can be shown to be also
decaying, i.e.,
$|\<\psi^-|\phi_{\mbox{\rm\scriptsize bg}}^+(t)\rangle|\rightarrow0$
for $t\rightarrow\infty$ for every $\psi^-\in\Phi_+$,
but it decays more slowly than the exponential~\cite{R20}.
Thus if one takes for $\phi^+$ the $|K^0\rangle$ state prepared e.g., by the
reaction $p\pi^-\rightarrow\Lambda K^0$ and for the observables
$|\psi^-\rangle\<\psi^-|$ projectors on the $\pi^+\pi^-$ space one obtains
according to the {\em exact} equation~(\ref{G19}) also a
term $|\<\pi^+\pi^-|\phi_{\mbox{\rm\scriptsize bg}}^+(t)\rangle|$.
This term vanishes more slowly than
to the rapidly disappearing
$\mbox{e}^{-i\Gamma_St}|\<\pi^+\pi^-|K^0_1\rangle|$.
This may provide an alternative mechanism to explain the $\pi\pi$
decay mode of the prepared $K^0$ long after the
$K_S^0=K_1^0$ has vanished.

The {\em third} step outside the space to obtain the basis of eigenvectors is
the {\em general complex basis vector} expansion. It includes in addition to the
ordinary Gamow kets~(\ref{G9}) also higher order Gamow kets~(\ref{G12}) which
occur
when (and if) the $S$-matrix has poles of order ${\cal N}>1$. Instead of writing
down the general expansion we restrict ourselves here to the special case that
there are no
bound states, there are two resonances at $z_{R_1}$ and $z_{R_2}$ and there is
one second order pole at $z_d=E_d-i\gamma_d$. Then the following basis system
expansion holds for $\phi_+\in\Phi_-$:
\begin{eqnarray}
\phi^+&=&
|z_d^-\rangle^{(0)}(-2\pi ia_{-2})^{(1)}\langle^+z_d|\phi^+\rangle-
|z_d^-\rangle^{(1)}(-2\pi ia_{-2})^{(0)}\langle^+z_d|\phi^+\rangle
\nonumber\\
&+&\sum_{i=1}^{2}
|z_{R_i}^-\rangle(-2\pi ia_{-1}^{(i)})\langle^+z_{R_i}|\phi^+\rangle
+\int_0^{-\infty_{II}}d\omega|\omega^+\rangle\langle^+\omega|\phi^+\rangle
\label{G22}
\end{eqnarray}
the $a_{-2}$, $a_{-1}^{(i)}$ are the expansion coefficients in the Laurent
series expansion of the $S$-matrix at the poles $z_d$, $z_{R_1}$ and $z_{R_2}$,
respectively.

The important distinction to~(\ref{G16}) is that this basis system contains
Jordan
vectors and the Hamiltonian is not diagonal but can only attain the Jordan
normal form:
\begin{equation}
\left(
\begin{array}{c}
\langle\psi^-|H^\times|z_d^-\rangle^{(0)}\\
\langle\psi^-|H^\times|z_d^-\rangle^{(1)}\\
\langle\psi^-|H^\times|z_{R_1}^-\rangle\\
\langle\psi^-|H^\times|z_{R_2}^-\rangle\\
\langle\psi^-|H^\times|\omega^+\rangle
\end{array}
\right)
=
\left(
\begin{array}{ccccc}
z_d&0\\
1&z_d\\
&&z_{R_1}\\
&&&z_{R_2}\\
&&&&(\omega)\\
\end{array}
\right)
\left(
\begin{array}{c}
\langle\psi^-|z_d^-\rangle^{(0)}\\
\langle\psi^-|z_d^-\rangle^{(1)}\\
\langle\psi^-|z_{R_1}^-\rangle\\
\langle\psi^-|z_{R_2}^-\rangle\\
\langle\psi^-|\omega^+\rangle
\end{array}
\right)
\label{G23}
\end{equation}
The time evolution of the basis vectors on the r.h.s.\ of~(\ref{G23})
is again given by the semigroup~(\ref{G8}), i.e., they have an arrow of time.
However, now in addition to the exponential dependence the time
evolution operator also
transforms according to~(\ref{ar22-1})
inside the two dimensional eigenspace ${\cal M}_{z_d}$ with an additional
linear time dependence. That second order poles of the $S$-matrix will introduce
an additional linear time dependence in the decay law has been known for
long time~\cite{R15}, only it was not clear what the vector was that evolved in
this way. This vector $\mid z_d^-\rangle^{(1)}$
has now been defined.
In addition the new result~(\ref{ar22-2})
shows
that the different values of $k$ get mixed up
by the time evolution.

Whereas there is no doubt that ordinary, $0$-order,
Gamow vectors will be the suitable vectors
to describe resonance states because of their properties
II.1\ldots II.5 above we have no idea what the physical meanining
of the higher order Gamow vectors may be, if any. In contrast to the fact that
ordinary Gamow states have been identified in abundance,
e.g., through their Breit-Wigner profile
in scattering experiments, or the exponential decay law~\cite{R5},
there is no convincing evidence for the existence of higher order poles
in nature~\cite{R6}.
The k-th order Gamow states and their time evolution~(\ref{ar22-2})
 are completely new and
unusual. Their effect should also be so overwhelming that the meager evidence
for higher order poles of the $S$-matrix which has been discussed in the past
($A_2$-splitting in particle
physics, $^8\mbox{Be}$ in nuclear physics) would not be able to account for it.
It is possible that there does not exist anything in nature that is described by
higher order Gamow vectors and first order resonance poles is all there is. But
since there is no theoretical reason against higher order poles of the
$S$-matrix and these higher order Gamow ``states'' emerge naturally from
the poles, it is worthwhile
to investigate their properties further~\cite{R7}.
 The only place that we can think to
look for effects of these higher order Gamow states are the high-multiplicity
events in high energy hadronic and nuclear collisions. That a quantum
mechanically rather pure initial state of two hadrons can result in a high
multiplicity event could have its origin in the highly impure ``resonance''
state
associated with the ${\cal N}$-dimensional subspace of higher order Gamow kets.

\section{Reversed time evolution and time reversal transformation}

An irreversible time evolution on the microphysical level immediately leads to
the question as to the time reversal transformation $A_T$. In the usual
reversible time evolution~(\ref{G6}) one always has with a state $W(t)$ also a
state $W(-t)$ (or with the state vector $\phi(t)$ also a state vector
$\phi(-t)=\mbox{e}^{-i(-2t)H}\phi(t)$).
The time
reversed state defined by $W^T(t)\equiv A_T^{-1}W(t)A_T$
or $\phi^T=A^{-1}_T\phi$
can therefore be identified with the negative time state:
\begin{equation}
W^\prime(-t)\equiv W^T(t)=W(-t);\,\,\,\,\phi^T(t)=\phi(-t)
\,\,\,\,\phi^\prime(-t)=\phi(-t)
\label{G25a}
\end{equation}
or in terms of the wave function since $A_T$ is antiunitary:
\begin{equation}
\phi^\prime({\bf x},-t)\equiv
\phi^T({\bf x},t)=
\phi^*({\bf x},t)
\label{G25b}
\end{equation}
For irreversible time
evolution one has the two semigroups~(\ref{G8+}) and~(\ref{G8}):
\begin{equation}
U^\dagger(t)|_{\Phi_-}\subset U_-^\times(t)
\hspace{10pt}
\mbox{in the space of states}
\hspace{10pt}
\Phi_-\subset
\Phi_-^\times
\hspace{10pt}
\mbox{for $t\leq0$}
\label{G24}
\end{equation}
and
\begin{equation}
U^\dagger(t)|_{\Phi_+}\subset U_+^\times(t)
\hspace{10pt}
\mbox{in the space of observables}
\hspace{10pt}
\Phi_+\subset
\Phi_+^\times
\hspace{10pt}
\mbox{for $t\geq0$}.
\label{G25}
\end{equation}
Therefore a state vector at the negative of the time $t$,
i.e.,
$\phi(-t)=\phi(|t|)$ cannot be obtained from $\phi(0)$ by this semigroup
transformation.
Thus there is in general no negative time {\em state} $W(-t)$ (or $\phi(-t)$)
which the time reversed state $W^T(t)$ (or $\phi^T=A_T^{(-1)}\phi$)
could be identified with. In particular one cannot
have the standard requirement~(\ref{G25a}), $A_T$ cannot be the operator that
transforms every conceivable state $W(t)$ into $W(-t)$. The operator
$U(|t|)|_{\Phi_-}$ is not continuous operator from $\Phi_-$ to $\Phi_-$
but
transforms out of the space of states $\Phi_-$ into the space $\Phi_+$.

A mathematically consistent resolution of the problem
with the time reversal operator, therefore,
would be to define
\begin{equation}
A_T:\Phi_-\rightarrow\Phi_+;
\hspace{5pt}
\Phi_+\ni\psi^-=A_T\phi^+,
\hspace{5pt}
\phi^+\in\Phi_-
\label{G27}
\end{equation}
This is indeed the solution suggested by conventional scattering theory where
the {\em in}-states $\phi^+$ or ($\phi^{\mbox{\rm\scriptsize in}}$)
are the time reversed of the so called {\em out}-states
$\psi^-$ or ($\psi^{\mbox{\rm\scriptsize out}}$).
(The ``{\em out}-states'' $\psi^{\mbox{\rm\scriptsize out}}$
are actually observables and not states because they are specified by the
detector whereas states are specified by the preparation apparatus
(accelerator)). This solution is based on the standard $A_T$ transformation
properties of the eigenkets of the exact Hamiltonian $H$
\begin{equation}
A_T|E^\pm,\eta\rangle
=\alpha
|E^\mp,\eta_T\rangle;
\,\,\,\,
A_T^2=(-1)^{2j}1,
\label{G28}
\end{equation}
which are defined by the Lipmann-Schwinger equation
\begin{equation}
|E^\pm,\eta\rangle=
|E,\eta\rangle +
\frac{1}{E-H\pm i\varepsilon}
V|E,\eta\rangle
\label{G29}
\end{equation}
($\eta$ are the degeneracy quantum numbers which include angular momentum (spin)
$j$ and $|E,\eta\rangle$ are
the eigenkets of $(H-V)=H_0$).

However~(\ref{G27}) would mean that $A_T$ transforms observables into states
(and
vice versa) and would therefore lead back to the identification of the set of
observables with
the set of states. Though $\Phi_+\cap\Phi_-\neq\emptyset$ (zero vector), which
means that there are vectors $\phi\in\Phi=\Phi_++\Phi_-$
which can represent states as well as observables, in general one cannot
postulate that every observable $|\psi\rangle\langle\psi|$ can be prepared as a
state. E.g., in a typical scattering experiment the ``{\em
out}-states'' represent highly correlated
spherical waves whereas the prepared {\em in}-states are typically two
uncorrelated plane waves (e.g., two colliding monochromatic beams). The time
reversal of this experiment would require a preparation apparatus that prepares
highly correlated (with fixed phase relationship) incoming
spherical waves that would be
scattered into two uncorrelated plane waves. An apparatus that would accomplish
this is impossible (or highly improbable)
to build, at least in this world. Thus, not for every
preparable state $W$ can one also prepare a state which would be described by
its time reversal transformed $W^T=A_T^{-1}WA_T$
(for another example see, e.g.\ chapter~13 of ref.~\cite{R13}).
This means that neither of the two quantities equated in the standard theory
by~(\ref{G25a}) may have a physical meaning in
terms of a preparation
procedure.

The division of $\Phi$ into
$\Phi_-$
(for states) and $\Phi_+$ (for observables)
that we obtained from the arrow of time
is not contradicted by the physics
of time reversal (because
one can build a rotated, a translated or even a parity
transformed preparation apparatus but one cannot build a time reversal
transformed preparation apparatus). But it is just in contradiction
with the standard theoretical description~(\ref{G27}) for the time
reversal operator.
Therefore, if irreversible processes on the microphysical level are to be
described, we
need a time reversal operator more general than the one conventionally used in
non-relativistic quantum mechanics and relativistic field theory. Wigner has
already provided such a time reversal operator~\cite{R16} which has also been
mentioned a few times in the literature~\cite{R17},\cite{R18}. But so far only
the $A_T$ with the standard property has found acceptance in physics.

The time reversal operator $A_T$ is not defined by its action on states
like~(\ref{G25a}) and~(\ref{G25b}), but by its relation to the
observables~\cite{R10},~\cite{R18}.
In general, the quantum mechanical operator $A_T$ representing time reversal
\[
T
\left(
\begin{array}{c}
t\\
{\bf x}
\end{array}
\right)
=
\left(
\begin{array}{c}
-t\\
{\bf x}
\end{array}
\right)
=-gx
\hspace{15pt}
g=
\left(
\begin{array}{cccc}
1\\
&-1\\
&&-1\\
&&&-1
\end{array}
\right)
\]
is an element of the (co)representation~\cite{R16} of space-time
transformations.
Space-time transformations (i.e., the extended (by time reversal and space
reflection) Poincar\'e group for relativistic space time and the extended
Galilei group for non-relativistic space-time
$\left(\begin{array}{c}t\\{\bf x}\end{array}\right)$)
were represented by unitary (and antiunitary for $A_T$)
operators in the Hilbert space. The time reversal operator $A_T$
is therefore defined (not by its action on the states) by its relation to the
other symmetry operators like the space reflection $U_P$ and the restricted
space time transformations
$U\left(\left(\begin{array}{c}t\\{\bf x}\end{array}\right),\Lambda\right)$.
An example of such a relation is~\cite{R18}
\begin{equation}
A_TU\left(\left(\begin{array}{c}t\\{\bf x}\end{array}\right),\Lambda\right)
A_T^{-1}=
U\left(\left(\begin{array}{c}-t\\{\bf x}\end{array}\right),g\Lambda g\right)
\end{equation}
From this one obtains the relation of $A_T$ to the observables, which are
the generators of $U(x,\Lambda)$. Examples of these relations
are
\begin{equation}
A_TP_iA_T^{-1}=-P_i,
\hspace{15pt}
A_TJ_iA_T^{-1}=-J_i,
\hspace{15pt}
A_TU_PA_T^{-1}=\varepsilon_T\varepsilon_I U_P,
\label{G31}
\end{equation}
\begin{equation}
A_THA_T^{-1}=H,
\hspace{15pt}
A_TH_0A_T^{-1}=H_0,
\hspace{15pt}
A_TSA_T^{-1}=S.
\label{G32}
\end{equation}
In here the generators $P_i$, $H$, $J_i$ represent momentum, energy, angular
momentum, respectively. The $S$-operator is a complicated function of
the interaction Hamiltonian $V=H-H_0$
 and $U_P$ is the unitary and hermitian parity operator normalized to
$U_P^2=1$.
The quantities
\[
\varepsilon_T=A_T^2
\hspace{25pt}
\varepsilon_I=(U_PA_T)^2\equiv A_I^2
\]
are real phase factors which define the 4 different extensions of the restricted
space-time symmetry transformations by space inversion $P=g$, time inversion
$T=-g$ and space-time inversion $I=PT=-1$.
(At this level where we have not talked about any charges, $U_P$ could also be
interpreted as representing the usual $CP$). Of the 4 possible extensions
$(\varepsilon_T,\varepsilon_I)=(\pm1,\pm1)$ the almost exclusive
choice~\cite{R18},\cite{R10}  for
$(\varepsilon_T,\varepsilon_I)$ is:
\begin{equation}
\varepsilon_T=(-1)^{2j}
\hspace{20pt}
\varepsilon_I=(-1)^{2j}
\hspace{20pt}
\mbox{where $j$ is the spin}
\label{G33}
\end{equation}
With this choice the only possibility for $A_T$ is~(\ref{G27}) which in the
interpretation requires to identify the set of states with the set of
observables (i.e., no arrow of time) and to assign to every $W(t)$
a $W^T(t)\equiv A_T^{-1}W(t)A_T$ fulfilling~(\ref{G25a}). This is in
contradiction
to the experience that at least for some states it is highly improbable to also
prepare their time reversed states
(cf.~ remark above and ch.~13 ref.~\cite{R13}).

A way out would be to give up either
irreversible time evolution or the time reversal operator. But since time
reversal invariance, defined by~(\ref{G31}) and~(\ref{G32})
has consequences which can be tested
experimentally, e.g., reciprocity relations, it is useful to retain the notion
of $A_T$ also if one includes irreversible time evolution. We therefore want to
explore the three other possibilities for
$(\varepsilon_T,\varepsilon_I)$ which do not fulfill~(\ref{G33}), i.e., the
other extensions of the space time symmetry groups provided by
Wigner~\cite{R16}. All three unconventional extensions
involve {\em time-reversal doubling} of the representation spaces. This
will introduce a further label $r$ in addition to the quantum numbers which we
called $\eta$ in~(\ref{G29}).
For $\eta$ we will choose angular momentum (spin)~$j$, its component~$j_3$
and other intrinsic quantum numbers~$n$,
which we do not specify further: $\eta=j_3,j,n$.
Thus the basis vectors are denoted by $|E^\pm,j_3,j,n;r\rangle$.

The four possible cases, of which the standard case~(\ref{G33}) is given in the
first row, are listed in the following table.
\begin{center}
{\bf Table 1}: Extensions of the space-time symmetry groups by $P$ and
$T$\\[15pt]
\begin{tabular}{|cccc|}
\hline
\multicolumn{2}{|c}{Characterization of the}&&\\
\multicolumn{2}{|c}{$P$ and $T$ extensions}&Representation of&
Representation of\\
$\varepsilon_T$&$\varepsilon_I$&$U_P$&$A_T$\\
\hline
$(-1)^{2j}$&$(-1)^{2j}$&1&$C$\\[5pt]
$-(-1)^{2j}$&$(-1)^{2j}$&
$
\left(
\begin{array}{cc}
1&0\\
0&-1
\end{array}
\right)
$&
$
\left(
\begin{array}{cc}
0&C\\
-C&0
\end{array}
\right)
$\\[15pt]
$(-1)^{2j}$&$-(-1)^{2j}$&
$
\left(
\begin{array}{cc}
1&0\\
0&-1
\end{array}
\right)
$&
$
\left(
\begin{array}{cc}
0&C\\
C&0
\end{array}
\right)
$\\[15pt]
$-(-1)^{2j}$&$-(-1)^{2j}$&
$
\left(
\begin{array}{cc}
1&0\\
0&1
\end{array}
\right)
$&
$
\left(
\begin{array}{cc}
0&C\\
-C&0
\end{array}
\right)
$\\
\hline
\end{tabular}
\end{center}
In this table $C$ is the well known operator:
\begin{equation}
C|E,j_3,j,n;r\rangle=\alpha(r)(-1)^{j-j_3}
|E,-j_3,j,n;r\rangle=
\sum_{j_3'}
\alpha(r)|E,j_3',j,n;r\rangle
C^{(j)}_{j_3'j_3}
\end{equation}
where $\alpha(r)$ is a phase factor and the matrix $C^{(j)}_{\mu\nu}$
is given by
\begin{equation}
C^{(j)}_{\mu\nu}=
(-1)^{j-\mu}\delta_{\mu,-\nu}
\hspace{20pt}
(-j\leq\mu,\nu\leq+j)
\label{G35}
\end{equation}
The index $r$ (= + or -)
labels two subspaces ${\cal H}(r)$ in which all the other known observables $B$
are identical, i.e., $B$ and $U_g$, where $g$ are continuous space-time
transformations not containing $P$ and $T$, are given by
\begin{equation}
B=
\left(
\begin{array}{cc}
B&0\\
0&B
\end{array}
\right);
\hspace{15pt}
U_g=
\left(
\begin{array}{cc}
U_g&0\\
0&U_g
\end{array}
\right).
\label{G36}
\end{equation}
The index $r$ thus also labels the rows and columns of the operator matrices in
the Table~1 and in~(\ref{G36}).

In the conventional case~(\ref{G33}) the label $r$ is not needed and $A_T$ is
given by (ignoring all the unspecified quantum numbers $n$)
\begin{equation}
A_T\mid E,j_3,j\rangle=
(-1)^{j+j_3}\alpha^\prime \mid E,-j_3,j\rangle
\label{G37}
\end{equation}
which we also write (suppressing from now on the quantum numbers $j_3$, $j$)
\begin{equation}
A_T\mid E\rangle=
\alpha
\mid E\rangle.
\label{G37b}
\end{equation}
The exact eigenvectors $\mid E^{\pm}\rangle$ which
are related to the $\mid E\rangle$
by (the formal solution of) the Lippmann-Schwinger equation~(\ref{G29}), have
the standard $A^T$ transformation property~(\ref{G28})

In the conventional case~(\ref{G37}),~(\ref{G28}) we have {\em one} Hilbert
space
${\cal H}$, {\em one} RHS $\Phi=\Phi_++\Phi_-\subset{\cal H}\subset\Phi^\times$;
$\Phi_+\cap\Phi_-\neq\emptyset$
and {\em one} pair of RHS's of Hardy class type~(\ref{G4a}),~(\ref{G4b}).
 The operator $A_T$ can only
be defined as in~(\ref{G27}), i.e.:
\begin{equation}
A_T:\Phi_\pm\rightarrow\Phi_\mp;\,\,\,\,\,\,
A_T^\times:\Phi^\times_\pm\rightarrow\Phi^\times_\mp
\label{G38}
\end{equation}
which means that the two spaces $\Phi_-$ and $\Phi_+$ are $A_T$ transforms
of each other. In our earlier discussion of the scattering experiment we have
already concluded that this cannot be possible
for empirical reason. Thus, if one has a quantum
mechanical arrow of time, then the time reversal operator cannot be defined in
the standard way with $A_T^2=+1$ (or $A_T^2=+(-1)^{2j}$).

Of the three unconventional cases the second and the third line of Table~1 gives
the cases in which $A_T$ transforms between parity eigenspaces of opposite
(relative) parity. In these cases the label $r$ can be given by the relative
parity and is therefore also not needed. We therefore choose the case in the
fourth line of the Table~1 characterized by
($\varepsilon_T=-(-1)^{2j}$, $\varepsilon_I=-(-1)^{2j}$).
In this case the action of $A_T$ is given by
\begin{equation}
A_T\mid E,r\rangle=\alpha(r)\mid E,-r\rangle;\,\,\,\,
\alpha^*(r)\alpha(-r)=\varepsilon_T=(-1)(-1)^{2j}
\label{G39}
\end{equation}
and the action of $A_T$ upon the exact energy eigenvectors
$\mid E^{\pm},r\rangle$ is given by
\begin{equation}
A_T\mid E^{\pm},r\rangle=\alpha(r)\mid E^{\mp},-r\rangle
\label{G40}
\end{equation}
In this new case we have two RHS's labeled by the index $r$,
$\Phi^r\subset{\cal H}^r\subset\Phi^{r\times}$
and two pairs of the RHS's of Hardy classes, in place of the one
pair~(\ref{G4a}) and~(\ref{G4b}):
\begin{equation}
\Phi_+^r\subset{\cal H}\subset\Phi_+^{r\times}\,\,\,\,
\mbox{and}\,\,\,\,
\Phi_-^r\subset{\cal H}\subset\Phi_-^{r\times},\,\,\,
r=\pm
\label{G41}
\end{equation}
\begin{equation}
\mbox{for any $\phi^+\in\Phi_-^r$ we have a $\psi^-\equiv
A_T\phi^+\in\Phi_+^{-r}$}
\label{G42}
\end{equation}
\begin{equation}
\mbox{for any $\psi^-\in\Phi_+^r$ we have a $\phi^+\equiv
A_T\psi^-\in\Phi_-^{-r}$}
\label{G43}
\end{equation}
From this we conclude that the operator $A_T$ maps the space $\Phi^r_{\pm}$
(continuously, one to one and) onto the space $\Phi^{-r}_{\mp}$
\begin{equation}
A_T:\Phi^r_{\pm}\rightarrow\Phi^{-r}_{\mp}\,\,\,\,
r=+,-
\label{G44}
\end{equation}
The conjugate operator which is defined as the extension of the adjoint operator
$A_T^\dagger:{\cal H}^r\rightarrow{\cal H}^{-r}$
according to
\begin{equation}
A^\dagger_T|_{\Phi^r}\subset A^\dagger_T\subset A^\times_T
\,\,\,\,\mbox{in}\,\,\,\,
\Phi^{r}\subset{\cal H}^r\subset\Phi^{r\times},
\label{G45}
\end{equation}
is then a (continuous, one to one) mapping between the corresponding dual spaces
\begin{equation}
A_T^\times:
\Phi^{r\times}_\pm\rightarrow
\Phi^{-r\times}_\mp\,\,\,\,
r=+,-\,\,.
\label{G46}
\end{equation}

Thus
an operator $A_T$, which is compatible with our physical interpretation of the
spaces $\Phi_+$ and $\Phi_-$
has indeed been given by Wigner in~\cite{R16}
for the case $\varepsilon_T=\varepsilon_I=-1(=-(-1)^{2j})$.
In this case $A_T$ (and $A_T^\dagger$) transforms --- according to~(\ref{G44})
--- from the space $\Phi^r_+$ ($r=+$), which contains vectors representing
properties of the outgoing scattering products of our real experiment, into
the space $\Phi^{-r}_-$, which contains {\em in}-state vectors of scattering
experiment which we cannot prepare (e.g., incoming spherical waves with fixed
phase relations). Vice versa, the space $\Phi^r_-$ (containing vectors that
represent real preparable {\em in}-states) is mapped by $A_T$ onto $\Phi^{-r}_+$
(containing properties which we cannot observe).

The same arguments apply according to~(\ref{G46}) to the microphysical resonance
states. The exponentially decaying Gamow vector $\psi^G=\mid
z_R,r^-\rangle\in\Phi_+^{r\times}$, $z_R=E_R-i\Gamma/2$, is mapped into a
vector $\mid z_R^*,-r^+\rangle\in\Phi_+^{-r\times}$
which exponentially decreases into the {\em negative} time direction. And the
Gamow state of our resonance scattering experiment,$\tilde{\psi}^G=\mid
z_R^*,r^+\rangle\sqrt{2\pi\Gamma}\in\Phi_+^{r\times}$,
which exponentially grows from $t=-\infty$ to $t=0$ (the time when the
preparation is completed and the registration begins)
is mapped by $A_T^\times$ into a vector $\mid z_R,-r^-\rangle\in\Phi^{-r}_+$
which like the $\mid z_R^*,-r^+\rangle$ cannot be detected in our scattering
experiment.

Thus mathematically, due to the time reversal doubling, we have two arrows of
time pointing in opposite directions. For $r=+$ we have two
semigroups~(\ref{G8+}),~(\ref{G8})
both evolving into the same direction of time. For
$t\leq0$ we have the semigroup
$U_-^\times=e^{-iH^\times t}$
(of growth) and for $t\geq0$ we have the semigroup
$U_+^\times=e^{-iH^\times t}$
(of decay). These provide our arrow of time. The RHS's~(\ref{G41})
with $r=-$ describe the time-reversal image of our physical experiments; this
time-reversed experiment we will find impossible to prepare.

One can show that like in the conventional case also in this new case
with $(\varepsilon_T=-(-1)^j,\,\,\varepsilon_I=-(-1)^j)$
we have
\begin{equation}
\mid E,r^+\rangle=
\mid E,r^-\rangle S(E)=
\mid E,r^-\rangle e^{2\delta(E)}
\,\,\,\,\mbox{for}
\,\,\,\,r=\pm,
\label{G47}
\end{equation}
(where $\delta(E)$ is the phase shift and $S(E)$ the $S$-matrix).
This is the consequence of ``time reversal invariance'' defined
by~(\ref{G31}) and~(\ref{G32}).
This means that the two spaces $\Phi_-^r$ (describing states)
and the two spaces $\Phi_+^r$ (describing observables) with different values
of~$r$, $r=+$ and $r=-$, are not intermingled by the dynamics given by $H$ or
the
$S$-operator. The experimentally tested consequences of time reversal invariance
like reciprocity relation remain intact separately for each value of $r$.

In conclusion, we have seen that the quantum mechanical arrow of time and
irreversible time evolution on the microphysical level (as exemplified by all
quantum mechanical resonance states) are not in contradiction to time reversal
invariance as defined by~~(\ref{G31}) and~(\ref{G32}).
However, for quantum physical systems with
irreversible time evolutions (resonances) the time-reversal operator $A_T$
is not the standard operator with $A_T^2=(-1)^{2j}$.
The price that we have to pay for describing irreversible time evolution and
time reversal invariance in a consistent way is the doubling of the spaces. One
pair of spaces,~(\ref{G41}) with $r=+$, contains microphysical states that
became and decay in our time direction. The other,~(\ref{G41}) with $r=-$,
contains microstates that became and decay in the opposite time direction.
Time-reversal invariance, as defined by~(\ref{G31}) and~(\ref{G32})
for the observables, does
not lead to a time symmetry for the states, like~(\ref{G25a}) and~(\ref{G25b}).
This is in agreement with the empirical facts that some conceivable
time-reversed states are highly improbable and practically impossible to
prepare~\cite{R13}. Theoretically, the time symmetry of the observables given
by~(\ref{G31}) and~(\ref{G32})
can be broken for the states in two different ways leading to two
arrows of time, $r=+$ and $r=-$.
We belive that the principle, if any, that selects the one arrow
over the other lies outside the scope of the theory

\end{document}